\documentclass[10pt,aps,amsmath,showpacs,amssymb,nofootinbib,pre,twocolumn]{revtex4}
\usepackage[dvips]{graphicx}
\usepackage[dvips,usenames]{color}

\newcommand{\av}[1]{\langle {#1} \rangle}

\begin{document}

\title{Slow dynamics and rare-region effects in the contact process on
  weighted tree networks} 

\author{G\'eza \'Odor}
\affiliation{Research Centre for Natural Sciences, 
Hungarian Academy of Sciences, MTA TTK MFA, 
P. O. Box 49, H-1525 Budapest, Hungary}

\author{Romualdo Pastor-Satorras} 
\affiliation{Departament de F\'\i sica i Enginyeria Nuclear,
Universitat Polit\`ecnica de Catalunya, Campus Nord B4, 08034
Barcelona, Spain}

\date{\today}

\begin{abstract} 
  We show that generic, slow dynamics can occur in the contact process
  on complex networks with a tree-like structure and a superimposed
  weight pattern, in the absence of additional (non-topological)
  sources of quenched disorder. The slow dynamics is induced by
  rare-region effects occurring on correlated subspaces of vertices
  connected by large weight edges, and manifests in the form of a
  smeared phase transition. We conjecture that more sophisticated
  network motifs could be able to induce Griffiths phases, as a
  consequence of purely topological disorder.
\end{abstract}

\maketitle

\section{Introduction}

The science of complex networks has witnessed a veritable explosion of
activity in the last decade, having become a powerful tool to
represent, analyze and understand a myriad of natural and man-made
systems, characterized by heterogeneous topological structures
\cite{barabasi02,mendesbook}.  In parallel with studies focused on the
topology and function of these systems, a large research effort has
been devoted to the analysis of non-equilibrium dynamical processes
running on top of complex networks and on the effects that those
complex substrates can have on their temporal behavior and also on
possible phase transitions
\cite{dorogovtsev07:_critic_phenom,barratbook}.  Such effects are
particularly important in networks with a heterogeneous contact
pattern, as in scale-free (SF) networks \cite{Barabasi:1999}, in which
the degree distribution, defined as the probability that an element
(vertex or node) is connected to $k$ others (has degree $k$), exhibits
long tails as $P(k)\sim k^{-\gamma}$, with a certain degree exponent
usually in the range $2< \gamma \leq 3$ \cite{barabasi02,mendesbook}.

One of the simplest non-equilibrium processes studied on networks is
the contact process (CP) \cite{harris74,liggett1985ips}. In this
model, vertices can be either occupied or empty. Empty vertices become
occupied on contact with an occupied neighbor of degree $k_i$ with a
rate $\lambda/k_i$.  On the other hand, occupied vertices become empty
with a unitary rate.  On a regular lattice the CP experiences a
non-equilibrium phase transition at a critical point $\lambda_c$,
separating an absorbing phase from an active one
\cite{marro1999npt,odorbook,Henkel}, whose order parameter is the
density of occupied sites $\rho$ in the steady state.  Thus, for
$\lambda<\lambda_c$, an absorbing phase with $\rho=0$ is observed,
while for $\lambda>\lambda_c$ the system reaches an active phase, with
$\rho >0$ in the thermodynamic limit. Through a systematic analysis
relying on numerical simulations
\cite{Castellano:2006,Castellano:2008,PhysRevE.83.066113,FFCR11} and
theoretical approaches based on the heterogeneous mean-field (HMF)
theory \cite{Castellano:2006,Castellano:2008,boguna09:_langev} a
picture has emerged on the behavior of the CP on complex networks that
emphasizes the strong effects of the network heterogeneity.  In
particular, it has been shown that its absorbing phase transition
exhibits a nontrivial finite-size scaling \cite{cardy88}, depending
not only on the number of vertices $N$, but also on the degree
fluctuations of the network, measured by the second moment of the
degree distribution $\av{k^2} = \sum_k k^2 P(k)$.  This dependence
induces very strong corrections to scaling in SF networks, which, if
properly taken into account, can actually be observed in numerical
simulations \cite{PhysRevE.83.066113,FFCR11}.

The study of the CP, as well as other processes, both in and out of
equilibrium \cite{dorogovtsev07:_critic_phenom,barratbook}, has shown
the important effects that the disordered, heterogeneous topological
structure in a network can impose on the dynamical systems defined on
top of it.  The effects of network disorder have recently been
extended one step further, in a series of papers
\cite{PhysRevLett.105.128701,odor:172,Juhasz:2011fk} dealing with the
possibility of observing Griffiths phases (GP) and rare-region (RR)
phenomena \cite{Griffiths,Vojta} at the phase transition of the CP on
complex networks\footnote{See also Ref.~\cite{2012arXiv1204.6282B} for
  an investigation of GPs in the context of the
  superconductor-insulator transition on SF networks.}.  In case of
regular lattices it is well known that quenched disorder\footnote{When
  the disorder is annealed it acts as a noise and it is hence
  irrelevant for the universal scaling behavior of the phase
  transition.}  can strongly alter the behavior of a phase transition,
imposing an anomalously slow relaxation \cite{harrisquenched}. This is
the consequence of inhomogeneities that can create, in the
thermodynamic limit, RRs of characteristic size $l$, with probability
$P_{RR}(l)\propto\exp(-c l)$, in which the system can stay active for
exponentially long times $\tau(l)\propto\exp(b l)$ below the critical
point $\lambda_c$.  In systems with spatially uncorrelated disorder, a
sharp phase transition is preserved, although it occurs at a critical
point $\lambda_c$ larger than the corresponding to the clean system
$\lambda_c^0$. In the region $\lambda_c^0 < \lambda < \lambda_c$, a GP
develops with a slow algebraic density decay $\rho \propto t^{-c/b}$,
$c/b$ being a non-universal exponent, which can be understood by
non-perturbative methods
\cite{PhysRevLett.59.586,sethna88,PhysRevB.54.3328,PhysRevLett.69.534,Igloi2005277}.
However, if the inhomogeneities are correlated in a subspace with a
diverging diameter in dimensions $d_c^- < d_{RR} < d$, where $d$ is
the dimension of the system and $d_c^-$ is the lower critical
dimension, the transition becomes completely smeared. The smearing is
due to fact that RRs can become infinite (in the thermodynamic limit)
and undergo independent phase transitions, ordering at different
values of $\lambda$ \cite{Vojta}. This situation prevents the
development of global order, in such a way that the clean sharp
transition to the absorbing state in destroyed, and the associated
singularities become rounded. In this case, above the clean critical
point, the density remains finite in the long time limit and
approaches this value in a power-law form
\begin{equation} \label{smearedPL}
\rho(t)-\rho(\infty) \sim t^{-\zeta} \,
\end{equation}
where the exponent $\zeta$ is also non-universal.
Close to $\lambda_c^0$, however, the initial decay of the density
follows an stretched exponential form 
\begin{equation}
  \ln[\rho(t)]\propto t^{d_r/(d_r+z)} \ ,
  \label{eq:32}
\end{equation}
where $z$ is the dynamical exponent and $d_r$ is the dimension of the
uncorrelated subspace, as can be seen by applying optimal fluctuation
theory arguments \cite{Vojta}.

In the case of the CP on complex networks, it has recently been shown
that an intrinsic quenched disorder\footnote{Temporal GPs have been
  considered in Ref.~\cite{TGP}.}  defined by a varying control
parameter $\lambda_i$, can induce GPs and other RR effects on random
Erd\H os-R\'enyi networks \cite{erdos61} below (and at) the
percolation threshold
\cite{PhysRevLett.105.128701,odor:172,Juhasz:2011fk}. The authors in
\cite{PhysRevLett.105.128701,Juhasz:2011fk} also considered the issue
of the effects of pure topological disorder for a CP with constant
$\lambda$.  By means of theoretical arguments and numerical
simulations, they conjectured that RR effects should be relevant only
on networks with a finite topological dimension $D$, defined in terms
of the number of vertices at a topological distance $\ell$ from a
given source, $N(\ell) \sim \ell^D$.  Therefore, for random networks,
in which the small-world property \cite{watts98} holds, namely, where
$N(\ell) \sim \exp(\ell)$ and $D\to\infty$, RR effects and GPs should
be absent.

In this paper we reexamine the issue of rare-region effects and the
associated slow dynamics in the CP on SF small-world networks with
additional topological disorder beyond random connectivity.  In
particular, we consider the effect of two topological features which
have been shown to induce a strong slowing down in dynamical
processes, namely a tree-like structured network
\cite{PhysRevE.78.011114} and a weight pattern superimposed on it
\cite{dynam_in_weigh_networ}.  By means of extensive simulations, we
provide numerical evidence that, although GPs cannot actually be
observed, the loop-less tree-like structure together with an
heterogeneous weight pattern induces a smearing of the CP critical
phase transition and slow, power-law dynamics for finite times.  This
can be understood via RR effects triggered by topological correlations
in the weights. Our results open the path to consider more complex
topological motifs, such as a large clustering \cite{watts98} or a
hierarchical community structure \cite{Fortunato201075}, as generators
of purely topological Griffiths phases in non-equilibrium phase
transition on complex networks.

\section{The contact process on weighted trees}
\label{sec:cont-proc-weight}

Both a tree-like structure and the presence of a weight distribution
are by themselves capable of slowing down the temporal evolution of
dynamical processes on random networks
\cite{dynam_in_weigh_networ,PhysRevE.78.011114,Castellano05}. The
induced slowing down is due to the generation of topological traps
that capture dynamics and prevent the fast and wide range spreading
naturally expected in small-world networks. In tree networks, traps
are created by the lack of loops, which implies that there is a single
path between any two vertices: Once activity is deep in the leaves of
a subtree, it cannot reach other sections of the network until it
first finds the exit from that subtree \cite{PhysRevE.78.011114}. In
weighted networks, on the other hand, activity can become trapped in
correlated sets of adjacent vertices, joined by edges carrying a large
weight, which prevent the exploration of other regions of the network
\cite{dynam_in_weigh_networ}. We expect that the combination of those
slowing down ingredients can lead to the emergence of RR phenomena.

\subsection{Network models}
\label{sec:model-definition}

We consider the CP on weighted tree networks constructed following the
Barab\'asi-Albert (BA) model \cite{Barabasi:1999}.  The choice of this
model is motivated by the fact that it allows to construct tree
structures in a very simple way, in contrast with other standard
network generation models, e.g Ref.~\cite{ucmmodel}. The BA is a
growing network model in which, at each time step $s$, a new vertex
with $m$ edges is added to the network and connected to an existing
vertex $s'$ of degree $k_{s'}$ with probability $\Pi_{s \rightarrow
  s'} = k_{s'} /\sum_{s''<s} k_{s''}$. This process is iterated until
reaching the desired network size $N$. The resulting network has a SF
degree distribution $P(k) \simeq 2 m^2 k^{-3}$; additionally, fixing
$m=1$ leads to a strictly tree (loop-less) topology.

Binary (non-weighted) BA networks can be transformed into weighted
ones by assigning to every edge connecting vertices $i$ and $j$ a
symmetric weight $\omega_{ij}$. We have considered two different
schemes for weight assignment based on the topology of the network,
i.e. weights are not random but depend on properties of the vertices
they connect. The two strategies are defined as follows:

(i) \emph{Weighted BA tree I (WBAT-I):} Multiplicative weights depending
on the degree of the adjacent vertices, namely
\begin{equation}
  \label{eq:2}
  \omega_{ij} = \omega_0 (k_i k_j)^{-\nu},
\end{equation}
where $\omega_0$ is an arbitrary scale and $\nu$ is a characteristic
exponent with $\nu\geq0$.  With this selection, weights decrease
with increasing degree, in such a way that edges with the largest
weight connect vertices with lowest degree. Obviously, an unweighted
network will correspond to $\nu=0$.

(ii) \emph{Weighted BA tree II (WBAT-II):} Weights assigned according
to their age in the network construction, namely
\begin{equation}
  \label{eq:3}
  \omega_{ij} =\frac{|i-j|^x}{N},
\end{equation}
where the node numbers $i$ and $j$ correspond to the particular time
step in which they were first introduced in the network. Since the
degree of nodes decreases as $k_i\propto (N/i)^{1/2}$ during this
process, where $N$ is the size of the network, this selection with
$x>0$ favors connection between unlike nodes and suppresses
interactions between similar ones.

The presence of weights in the network affects the dynamics of the CP
in the rate at which empty vertices become occupied. Thus, the rate at
which an empty vertex $i$ becomes occupied on contact with an occupied
vertex $j$ is now proportional to $\lambda \omega_{ij}$. The
proportionality of this rate with $\omega_{ij}$ implies that the
creation of new particles takes place with larger probability between
vertices joined by a large weight edge. We can thus conclude that
activity can in principle become trapped in isolated connected subsets
of vertices, joined by large weight edges, subsets which are at the
same time connected among them only through small weight edges, which
transport activity only with very small probability. As we will see
below, these correlated subsets of vertices will play the role on rare
regions in the analysis of the CP in tree weighted networks.

\subsection{Heterogeneous mean-field analysis}
\label{sec:heter-mean-field}

An analytical understanding of the CP (and in general of any dynamical
process) on complex heterogeneous networks can be gained through the
application of heterogeneous mean-field (HMF) theory
\cite{pv01a,dorogovtsev07:_critic_phenom,barratbook}.  Thus, in the
case of symmetric weights depending on the degree at the edge's end
points, i.e. $\omega_{ij} = g(k_i, k_j)$, we can write, following
Ref.~\cite{dynam_in_weigh_networ}, the rate equation for the density
of occupied vertices of degree $k$, $\rho_k(t)$, namely
\begin{eqnarray}
  \label{eq:4}
  \dot{\rho}_k(t) &=& - \rho_k(t) + \lambda k [1-\rho_k(t)]  \times
  \nonumber \\
  &\times& \sum_{k'}\frac{g(k',k) P(k'|k)}{k' \sum_q g(k',q) P(q|k')}
  \rho_{k'}(t), 
\end{eqnarray}
where $P(k'|k)$ is the conditional probability that vertex $k$ is
connected to vertex $k'$ \cite{alexei}. The analysis of
Eq.~\eqref{eq:4} for general functions $g(k',k)$ and $P(k'|k)$
presents notable mathematical difficulties. We will then focus in the
case of uncorrelated networks, with $P(k'|k) = k' P(k') / \av{k}$
\cite{mendesbook} and multiplicative weights $g(k',k) = g(k)g(k')$
\cite{dynam_in_weigh_networ}, which correspond to the WBAT-I model,
with $g(k) = \sqrt{\omega_0} k^{-\nu}$. The former approximation is in
this case essentially correct, since BA networks are very weakly
correlated \cite{barrat2005rea}. We are thus lead to the simplified
rate equation
\begin{equation}
  \label{eq:1}
 \dot{\rho}_k(t) = - \rho_k(t) + \lambda  [1-\rho_k(t)] \frac{k^{1-\nu}
   \rho(t)}{\av{k^{1-\nu}}},
\end{equation}
where $\rho(t) = \sum_k P(k) \rho_k(t)$ is the total density of
occupied nodes in the network.

From Eq.~(\ref{eq:1}), we see that $\rho_k=0$ is always a solution.
The conditions for the presence of non-zero steady states can be
obtained by performing a linear stability analysis~\cite{marian1}.
Neglecting higher order terms, Eq.~(\ref{eq:1}) can be linearized as
$\dot{\rho}_k(t) \simeq \sum_{k'} L_{k k'} \rho_{k'}(t)$, with
\begin{equation}
  L_{k k'}  = -\delta_{k, k'} + \lambda  
    \frac{k^{1-\nu}P(k')}{\av{k^{1-\nu}}}.
\end{equation}
It is easy to see that the Jacobian matrix $L_{k k'}$ has a
  largest eigenvalue $\Lambda = \lambda-1$, associated to the
  eigenvector $v_k =k^{1-\nu}$. A nonzero steady state is only
possible for $\Lambda>0$, which translates in a critical threshold
\begin{equation}
  \lambda_c=1,
\end{equation}
independent of the degree distribution and the weight pattern of the
network\footnote{We do not expect, however, to obtain this threshold
  in numerical simulations, since it is a non-universal parameter
  \cite{Castellano:2006,FFCR11}.}.

More information on the active phase can be obtained by imposing the
steady state condition $\dot{\rho}_k(t)=0$, yielding the nonzero
solution
\begin{equation}
  \label{eq:5}
  \rho_k = \frac{\lambda k^{1-\nu} \rho/\av{k^{1-\nu}}}{1+\lambda
    k^{1-\nu} \rho/\av{k^{1-\nu}}}. 
\end{equation}
Inserting this expression on the relation $\rho=\sum_k P(k)\rho_k$, we
can obtain a self-consistent equation for $\rho$ in the continuous
degree approximation, substituting sums by integrals and approximating
the degree distribution by $P(k)= 2m^2 k^{-3}$ with $k \in [m,
\infty]$ for BA networks. The form of this approximation depends on
the value of $\nu$ (see Appendix for a detailed calculation). So, for
$\nu=0$, which corresponds to a an  unweighted BA network, we obtain a
particle density, close to the critical point, given by
\begin{equation}
  \label{eq:19}
  \rho \sim - \dfrac{\lambda-1}{\ln\left(\frac{\lambda-1}{\lambda} \right)}.
\end{equation}
We thus have a critical point $\lambda_c=1$ and a mean-field
exponent $\beta=1$, with additional weak logarithmic corrections. For
$\nu>0$, on the other hand, we obtain, close to the critical point
\begin{equation}
  \label{eq:6}
  \rho \sim \lambda -1,
\end{equation}
recovering the homogeneous MF result exponent $\beta=1$.

In order to find the decay of the order parameter at the critical
point, we look at the time evolution of $\rho(t)$, setting
$\lambda=1$, namely
\begin{equation}
  \label{eq:20}
  \dot{\rho}(t) = \sum_k P(k) \dot{\rho}_k(t) = -
  \frac{\rho(t)}{\av{k}} \sum_k P(k) k \rho_k(t).
\end{equation}
Performing a quasi-static approximation
\cite{michelediffusion,boguna09:_langev}, substituting $\rho_k(t)$ by
the form given by Eq.~\eqref{eq:5} with $\lambda=1$, and performing
again a continuous degree approximation, changing sums by integrals,
we obtain a simple differential equation that can be easily
solve. Thus, for the case $\nu=0$ we are led to (see Appendix)
\begin{equation}
  \label{eq:23}
  \rho(t) \sim [t \ln(t)]^{-1},
\end{equation}
which recovers the homogeneous mean-field result, $\rho(t) \sim
t^{-\alpha}$, with $\alpha=1$ plus additional logarithmic
corrections. For $\nu>0$, we obtain instead a full homogeneous
mean-field result, $\alpha=1$.

We conclude from this analysis that the CP on the weighted BA model
should (at least at mean field level) show a critical behavior
compatible with HMF exponent in the thermodynamic limit for $\nu>0$,
while weak logarithmic correction are expected for $\nu=0$.

\subsection{Finite-size effects}
\label{sec:finite-size-scaling}

Recently in has been called attention on the finite-size effects in
the CP on heterogeneous networks, which can drive the transition so
strongly as to overcome the thermodynamic limit behavior
\cite{Castellano:2008,boguna09:_langev,FFCR11}. It is easy to check
that in the present case, these finite-size effects are very weak for
$\nu=0$ or irrelevant for $\nu>0$.  In fact, following in a simplified
form the reasoning in Ref.~\cite{boguna09:_langev}, we can write the
time evolution of the total density $\rho(t)$ as
\begin{eqnarray}
  \lefteqn{\dot{\rho}(t) =\sum_k \dot{\rho}_k(t) P(k)}  \nonumber \\
  &&=  \Delta \rho(t) -
  \frac{\lambda^2 \rho(t)^2}{\av{k^{1-\nu}}^2}  
    \sum_k \frac{k^{2(1-\nu)} P(k)}{[ 1+ \lambda
        k^{1-\nu} \rho(t) / \av{k^{1-\nu}}]},  \label{eq:24}
\end{eqnarray}
where $\Delta= \lambda-1$ and we have used
Eq.~\eqref{eq:5}. In the limit of very small particle
density in a network of finite size $N$, when $\rho \ll
\av{k^{1-\nu}}/ k_c^{1-\nu}$, where $k_c$ is the maximum degree of
cut-off in the network \cite{Dorogovtsev:2002}, we can approximate
\begin{eqnarray}
  \dot{\rho}(t) &\simeq& \Delta \rho(t) -
  \frac{\lambda^2 \rho(t)^2}{\av{k^{1-\nu}}^2}  
    \sum_k k^{2(1-\nu)} P(k)  \nonumber \\
    &=&\rho(t) \left[ \Delta - \lambda^2 \rho(t) g_\nu \right],
    \label{eq:13} 
\end{eqnarray}
where we have defined the parameter
\begin{equation}
  \label{eq:25}
  g_\nu = \frac{{\av{k^{2(1-\nu)}}}}{\av{k^{1-\nu}}^2} \sim \left\{
    \begin{array}{lcl}
      \ln (N) & \mathrm{for} & \nu=0  \\
      \mathrm{const} &\mathrm{for} & \nu>0  
    \end{array}\right. .
\end{equation}
The only nonzero solution of Eq.~\eqref{eq:13} gives thus the finite
size behavior
\begin{equation}
  \label{eq:33}
  \rho \simeq \frac{\lambda-1}{\lambda^3 g_\nu}.
\end{equation}
Form Eq.~\eqref{eq:25}, we can see that $g_\nu$ is size independent
for $\nu>0$, while it depends weakly (logarithmically) on $N$ for
$\nu=0$. We thus conclude that, even at the finite-size regime, the CP
in weighted BA networks is fully described by the homogeneous
mean-field critical point, with no strong corrections to scaling
except for the presence of weak, logarithmic ones at $\nu=0$.

Size effects also appear in the density decay with time at the
critical point. We can estimate those effects by considering the time
evolution of $\rho(t)$ at $\lambda=1$ ($\Delta=0$) in a finite
network. Again in the limit $\rho \ll \av{k^{1-\nu}}/ k_c^{1-\nu}$, we
can simply integrate Eq.~\eqref{eq:13}, to obtain the finite size
density decay in the limit of large times as
\begin{equation}
  \label{eq:30}
  \rho(t) \simeq [g_\nu t]^{-1}.
\end{equation}
Again we observe that density decay in time is given according to the
homogeneous mean-field theory for $\nu>0$, with logarithmic size
corrections at $\nu=0$.

To conclude this Section, we emphasize that the HMF solution presented
here is expected to be correct only in the case of non-tree looped
networks, since HMF theory is know to provide inaccurate predictions
in tree topologies
\cite{Castellano05,nohkim,dallasta_ng_nets,PhysRevE.78.011114}, and to
be accurate on weighted networks only within certain limits
\cite{dynam_in_weigh_networ,2010arXiv1011.2395B}. It is nevertheless
instructive to consider its results as a first order
approximation. Additionally, we must note that no HMF prediction is
possible for the WBAT-II, due to the additive nature of the weight
pattern.

\section{Numerical simulations}
\label{sec:numer-simul}

In a practical numerical implementation of the CP, the neighbors of
all nodes are stored in a dynamically growing table of vectors, in
such a way that networks up to size $N=8\times 10^7$ can be stored in
$7.2$ GB of computer memory. We follow the density decay of active
sites by starting from fully occupied networks. Active nodes are
selected randomly and deactivated with probability $1/(1+\lambda)$; otherwise 
in a randomly selected neighborhood a site activation occurs with
probability $[\lambda/(1+\lambda)] \omega_{ij}$.  Time is updated
by 1 Monte Carlo (MC) step after $N_a$ attempts, where $N_a$ denotes
the number of active nodes in the previous time step. The simulations
were run up to $t= 2^{22}$ - $2^{26}$ MC time steps and have been
averaged over $10^2$ to $10^6$ independent networks of different size.
By the statistical sampling we used master-worker type of
parallelization (MPI-C).  This has been been repeated for several
values of the control parameters: $\lambda$, $\nu$ and $x$.

\subsection{Unweighted looped BA networks}
\label{sec:ba-looped}

\begin{figure}[t]
\includegraphics[height=6 cm]{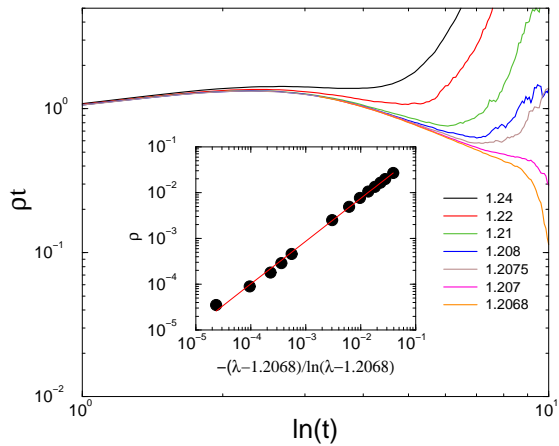}
\caption{\label{BAloopedcp} (Color online)
Density decay ($t\rho(t)$) as a function of $\ln(t)$ for the 
CP on unweighted looped BA networks with $m=3$ of size 
$N=8\times 10^7$. The different curves correspond to 
$\lambda=1.2068, ..., 1.24$ (from bottom to top). Inset:
Steady state density, showing agreement with HMF 
theory scaling. The full line shows a power-law fitting to the data 
points in the form $-0.36(5) x^{0.98(2)}$.}
\end{figure}

We fist consider the behavior of the CP on unweighted looped BA
networks, generated with $m=3$. We find a critical point around
$\lambda_c=1.2068(1)$ and a critical exponent $\beta=0.98(2)$ with
logarithmic corrections, as expected from Eq.~\eqref{eq:19}, see inset
in Fig.~\ref{BAloopedcp}.  The main plot of Fig.~\ref{BAloopedcp}
presents the particle density decay for different values of $\lambda$
around the estimated critical point, rescaled according to the HMF
prediction $\rho(t) \sim 1/[t \ln(t)]$. This prediction is fairly well
fulfilled by the simulations, as we can see from the almost straight
linear part in the $\ln[t\rho(t)]$ vs.  $\ln[\ln(t)]$ curve for
$\lambda=1.2068$, although finite size corrections are noticeable.  A
linear fit in the region $4 < \ln(\ln(t)) < 7$ for this value of
$\lambda$ results in the slope $-1.1(1)$, compatible with the
  HMF prediction.

The behavior observed unweighted looped networks is therefore
compatible with an standard critical point, reasonably described by
HMF theory, showing a clear supercritical phase reaching a stable
plateau of activity for $\lambda > \lambda_c$, an exponential activity
decay for $\lambda < \lambda_c$, and no apparent continuously varying
power-law decays of activity in the vicinity of the critical point, in
agreement with the results reported in Ref.~\cite{FFCR11}.

\subsection{Unweighted BA trees}
\label{sec:ba-trees}

We now investigate the effect of a loop-less tree structure, again in
the absence of weights.  For small network sizes, $N \leq 10^6$,
numerical results do not yield to simple interpretation.  S-shaped
density decay curves can be seen on log-log plots, in such a way that
the location of the transition point cannot be clearly
determined. This observation suggests the presence of corrections to
scaling that could be larger than those predicted in the simple HMF
analysis presented in Sec.~\ref{sec:nu0}. In fact, as we increase the
network size, a phase transition with power-laws (PL) seems to emerge,
but the corrections became negligible only for network sizes $N >
10^7$ nodes.
\begin{figure}[t]
\includegraphics[height=6 cm]{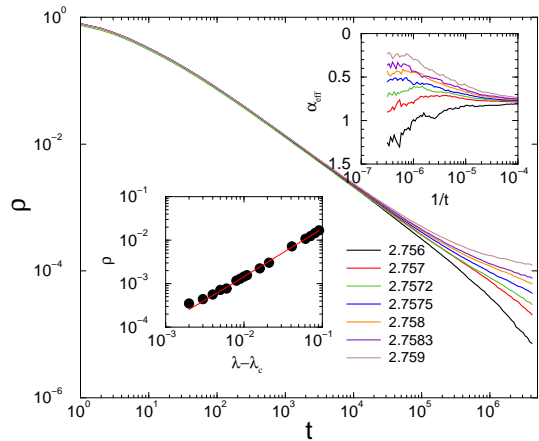}
\caption{\label{BAcp}(Color online)
  Density decay as a function of time for the CP on
  unweighted tree BA networks of size $N=8\times 10^7$. The different
  plots correspond to $\lambda=2.756, ..., 2.759$ (from bottom to
  top). Right inset: The corresponding effective exponents from a
  local slopes analysis (increasing $\lambda=$ from bottom to top). 
  Left inset: Steady state density with
  power-law fitting of the form (\ref{ssr}) for parameters 
  in opposite order (from top to bottom) as in the main graph.}
\end{figure}
In Fig.~\ref{BAcp} we show the particle density decay on networks of size
$N=8\times 10^7$ in a narrow scaling region around the critical point
given by $2.756 \leq \lambda \leq 2.759$. We have done a careful
analysis of the steady-state, with $\rho$ leveling off to constant
values, taking into account the effects of finite sizes.  By assuming
a critical transition (see left inset in Fig.~\ref{BAcp}) in the form
\begin{equation}\label{ssr}
\rho(\infty) = A |\lambda-\lambda_c|^{\beta} \ ,
\end{equation} 
we obtain a critical point $\lambda_c =2.757(1)$ with $A=0.287$ and
$\beta=1.20(5)$ by means of a least-squares fitting performed
discarding the smallest values of $\lambda$.  Note that this exponent
$\beta\simeq1.2$ is larger than the expected HMF result, an
observation hinting towards a failure of the HMF theory for the CP on
tree-like networks \cite{PhysRevE.78.011114}. The main plot of
Fig.~\ref{BAcp} depicts the decay in time of the particle density
$\rho(t)$ for different values of $\lambda$ around the estimated
critical point. The curves shown exhibit a decay exponent apparently
varying with the value of $\lambda$ at very large times. In order to
explore in more detail the decay of the density functions, we have
computed an effective decay exponents, defined as the local slope of
$\rho(t)$ as given by
\begin{equation}  \label{aeff}
  \alpha_{\rm eff}(t) = - \frac {\ln[\rho(t)/\rho(t')]} 
  {\ln(t/t^{\prime})} \ ,
\end{equation}
where $t$ and $t^{\prime}$ have been chosen in such a way that the
discrete approximate of the derivative is sufficiently smooth, see
right inset in Fig.~\ref{BAcp}. From this figure, we observe that the
decay exponents are indeed different, smaller than the HMF
prediction, which would lead to an effective exponent larger than one
in a log-log plot.  Moreover, the exponents thus computed are not well
defined for all values of $\lambda$, since the plots of $\alpha_{\rm
  eff}(t)$ do not show in general well-defined plateaus except for
very small regions at very large times. 

The conclusion we draw from the analysis of non-weighted BA trees is
that these network structures do not show evidence for the presence of
explicit RR effects, but are compatible with a standard
non-equilibrium phase transition. Evidences for RR effects, and in
particular for a smeared transition, are however compelling in the
case of weighted BA trees, as we will see below.

\subsection{Weighted BA trees with multiplicative weights: WBAT-I
  model}
\label{sec:weighted-ba-trees}

We next investigate the possibility of inducing RR effects in the
activity propagation of the CP by adding to the BA tree networks a
weight structure wich will act against the ``stronger gets stronger''
behavior of BA models by suppressing the dominancy of strongly
connected nodes.
 
In the first place, we consider the multiplicative weighting strategy
defined by model WBAT-I. In this case, we have as a theoretical
guideline for the behavior of the CP the results from the HMF analysis
presented in Sec.~\ref{sec:heter-mean-field}, which suggest a fully
homogeneous mean-field behavior, with $\beta=\alpha=1$. We note that
departures from this prediction are however not unexpected, given the
tree-like and weighted structure of the network model under
consideration.

\begin{figure}[t]
\includegraphics[height=6cm]{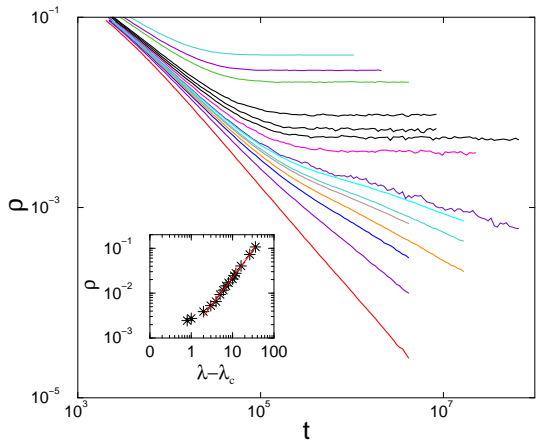}
\caption{\label{wbacp} (Color online)
  Density decay as a function of time for the CP
  on weighted BA trees generated with the WBAT-I model with exponent
  $\nu=1.5$. Network size $N=10^5$. Different curves correspond to
  $\lambda=$ 160, 156, 154, 149, 148, 147, 146, 145, 144.7, 144.2, 
144, 143.5, 143, 142, 140  (from top to bottom). 
Inset: Steady state density.}
\end{figure} 

For a small weight exponent $\nu=1/2$ we find a phase transition at
$\lambda_c=8.62(1)$, characterized by the mean-field type of density
decay: $\rho(t) \propto 1/t$ (not shown) \cite{karsai:036116}.  For
larger values of $\nu$, however, generic, continuosly changing
power-laws can be observed. In particular for $\nu=1.5$ simulations
performed on networks of size $N=10^5$ suggest the presence of a GP as
shown on Fig.~\ref{wbacp}.  Considering the steady state value of the
density, we can observe a very smooth approach to zero.  A power-law
fitting to the form $\rho = A |\lambda-\lambda_c|^\beta$ results is
$\lambda_c=143(1)$ and $\beta=1.3(1)$, as shown in the inset of
Fig.~\ref{wbacp}.  The value of $\beta$ obtained in this fitting is
again in disagreement with the HMF prediction from
Sec.~\ref{sec:heter-mean-field}.

\begin{figure}
\includegraphics[height=6cm]{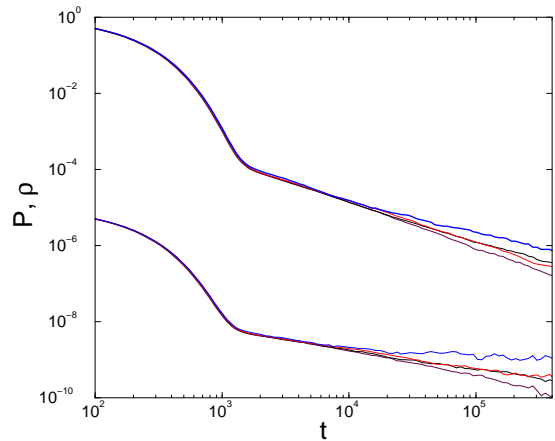}
\caption{\label{wbacp-seed} (Color online)
  Density decay as a function of time
  $\rho(t)$ (lowest branch curves) and survival probability $P(t)$
  (highest branch curves) for seed simulations of the CP on weighted
  BA trees generated with the WBAT-I model with exponent
  $\nu=1.5$. Curves correspond to $\lambda=145, 144, 143, 140$ (top to
  bottom for each branch). Network size $N=10^5$.}
\end{figure}

We have also performed spreading simulations, starting from a very
small initial seed of active sites \cite{odorbook}. These simulations
were performed on networks with $N=10^5$ nodes up to $t=4\times 10^5$
MC time steps.  Besides the density $\rho(t)$, we measured the
survival probability $P(t)$ \cite{marro1999npt}, averaged over $10^8$
independent runs on $10^3$ different network configurations.  As
Fig.~\ref{wbacp-seed} shows, not only the density, but also the
survival probability $P(t)$ curves exhibit slow, power-like behavior
for times larger than $t>2000$.  In the initial time region $t<1000$,
both $P(t)$ and $\rho(t)$ decay exponentially.

\begin{figure}[t]
\includegraphics[height=6cm]{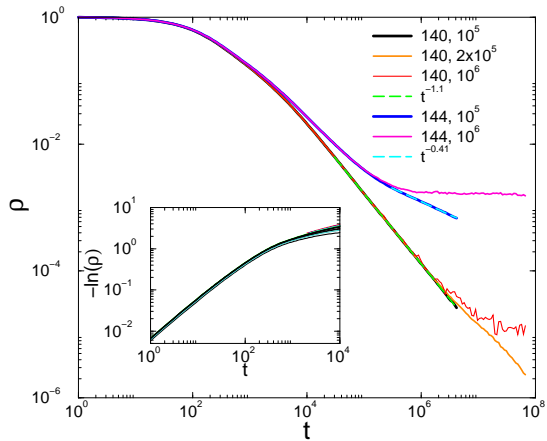}
\caption{\label{fsswbacp} (Color online)
  Density decay as a function of time
  $\rho(t)$ for the CP on weighted BA trees with a multiplicative
  weighting scheme (WBAT-I) with exponent $\nu=1.5$. Plots correspond
  to two sets of $\lambda$ (upper branch: $\lambda=144$, lower branch  
  $\lambda=140$) at different network sizes $N$.  Dashed lines
  represent PL fittings. Inset: Initial time region of the same data,
  showing an stretched exponential behavior near the transition.}
\end{figure}

Simulations performed for larger network sizes suggest, however, that
the apparent PLs observed in fact disappear in the thermodynamic limit
of large network size. In Fig.~\ref{fsswbacp} we plot the particle
density decay as a function of time for $\lambda =140$ and $144$
corresponding to increasing network sizes from $N=10^5$ up to
$N=10^6$. As can be seen from this plot, the apparent power laws,
fitted by dashed lines, that can be observed for $N<10^5$, turn into
saturation plateaus for $N=10^6$ at very late times.

This observation suggests that what we are actually observing in the
WBAT-I model is the tail behavior of a smeared transition
\cite{Vojta}, where the density has a constant nonzero value
$\rho_{st}=\rho(\infty)$ in the long time limit and $\rho(t)$
approaches this in a PL manner, see Eq.~\eqref{smearedPL}.  Additional
evidence of this picture come from the analysis of the initial time
decay of the particle density, with can be fitted to an stretched
exponential of the form $\rho\propto\exp(t^{-0.91})$ (see inset on
Fig.~\ref{fsswbacp}). This behavior is very close to an exponential
decay. However, comparing with the CP with correlated disorder on
regular lattices, at intermediate times of the smeared transition we
should observe a stretched exponential of the form in
Eq.~\eqref{eq:32}.  A value $d_r \gg z$ in Eq.~\eqref{eq:32} would fit
to the observed behavior, an expectation that is quite reasonable in a
random network with infinite topological dimensions.

The conclusion extracted from the analysis of this model is that the
presence of a smeared phase transition is quite probable. A smeared
transition can explain the saturation in the thermodynamic
$N\to\infty$ limit, due to the presence of correlated subspaces of
connected vertices of small degree, and correspondingly joined by
relatively large weights. These subspaces play the role of RRs, that
can remain active, leading to an initial stretched exponential
dynamics crossing over to power-law for longer times \cite{DV05}. 

In order to visualize the presence of these RRs, we have performed a
percolation analysis
\cite{stauffer94,ostojic06,1742-5468-2012-02-P02008} of our WBAT-I
networks. We consider a network of a given size $N$, and delete all
the edges with a weight smaller than a threshold
$\omega_\mathrm{th}$. For small values of $\omega_\mathrm{th}$, many
edges remain in the system, and they form a connected network with a
single cluster encompassing almost all the vertices in the
network. When increasing the value of $\omega_\mathrm{th}$, the
network breaks down into smaller subnetworks of connected edges,
joined by weights larger than $\omega_\mathrm{th}$. The expectation in
the latter case from standard percolation in networks is to observe
the presence of a largest cluster (the giant component) with a size
scaling with $N$, while the rest of the components should have a size
scaling logarithmically with $N$ at most \cite{Newman2010}.
\begin{figure}[t]
  \includegraphics[height=5cm]{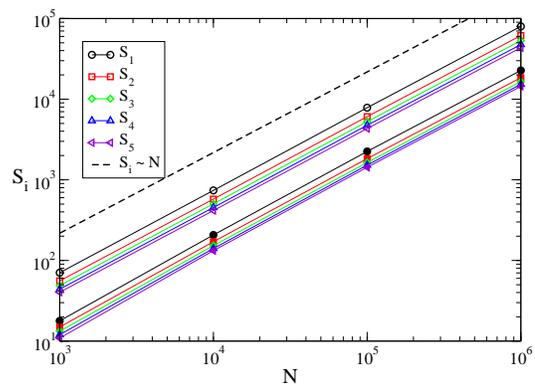}
  \caption{\label{perco} (Color online)
    Size $S_i$ of the $5$ largest clusters in a
    percolation analysis of the WBAT-I model with $\nu=1.5$ for
    $\omega_\mathrm{th} = 100 \omega_\mathrm{min}$ (hollow symbols)
    and $\omega_\mathrm{th} = 1000 \omega_\mathrm{min}$ (full
    symbols), where $ \omega_\mathrm{min}$ is the minimum weight in
    the network. The size of all components grows linearly with
    network size $N$, and is therefore infinite in the thermodynamic
    limit.}
\end{figure}
In Fig.~\ref{perco} we plot the average size (measured as the number
of vertices) of the $5$ largest components corresponding to a given
threshold $\omega_\mathrm{th}$. For the values of $\omega_\mathrm{th}$
considered, the networks break down in a number of
components. However, the size of the largest ones grows linearly with
the network size $N$, at odds with the expectations from a standard
percolation transition. These clusters, which can become arbitrarily
large in the thermodynamic limit, play the role of correlated RRs,
sustaining independently activity and smearing down the phase
transition.

This sort of analys allows also to explain why looped weighted
networks do not exhibit RR effects. In this case (data not shown), the
largest component scales again linearly with the network size, but the
next largest components scale in size logarithmically. They thus
become irrelevant in the thermodynamic limit and cannot sustain
activity at very large times.

\subsection{Weighted BA trees with age-dependent weights: WBAT-II model}

We finally consider the behavior of the CP on weighted tree BA
networks constructed with the age-dependent weighting scheme, defined
by the model WBAT-II. In this case, the weights are anti-correlated
with the degree, in opposition with the previous WBAT-I, in which
large degree vertices had small weight, with small degree vertices had
large weight. Here instead, weight is larger in edges connecting
vertices with different degrees, while edges between vertices of
similar degree have a small weight. 

\begin{figure}
  \includegraphics[height=6cm]{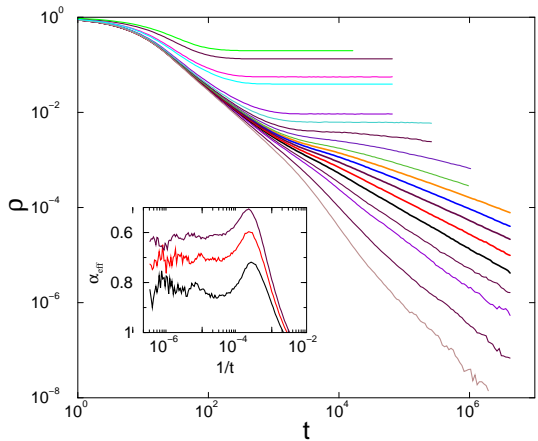}
  \caption{\label{gwbacp} (Color online)
    Density decay as a function of time
    $\rho(t)$ for the CP on weighted BA trees with a age-dependent
    weighting scheme (WBAT-II) with exponent $x=2$. Network size
    $N=10^5$. Different curves correspond to $\lambda=$ 6.75, 6.8, 6.85,
    6.87, 6.9, 6.92, 6.94, 6.96, 6.98, 7, 7.04, 7.1, 7,2, 7.4, 8.5, 9,
    12, 15 (from top to bottom). Inset: Corresponding local slopes for
    $\lambda=6.9, 6.92, 6.94$ (from bottom to top).} \end{figure}

Selecting an exponent for the weight strength $x=2$ and not very large
network sizes, $N=10^5$, generic power-laws can be observed for a
large range of $\lambda$ of values, as shown on Fig.~\ref{gwbacp},
which exhibit a very strong similarity with those found in the
multiplicative WBAT-I model, see Fig.~\ref{wbacp}. 
An analysis of the steady state density at large times in the region
$6.8 <\lambda < 9$ can be fitted again to the canonical form $\rho =
A |\lambda-\lambda_c|^\beta$, with $\beta=1.35(5)$, $A=0.02(5)$ and
$\lambda_c = 6.80(2)$. These values are
again in discrepancy with HMF expectation, but this fact is
altogether not surprising, since we do not have a proper HMF solution
for this model, and we can only proceed to compare with the
homogeneous result observed for the  WBAT-I model. The inset in
Fig.~\ref{gwbacp} shows the effective exponents $\alpha_{\rm eff}(t)$
as computed using the formula in Eq.~\eqref{aeff}. From the figure we
observe the presence of reasonably large plateaus that confirm the
presence of well-defined PLs with exponent varying with the value of
$\lambda$.

\begin{figure}[h,t]
  \includegraphics[height=6cm]{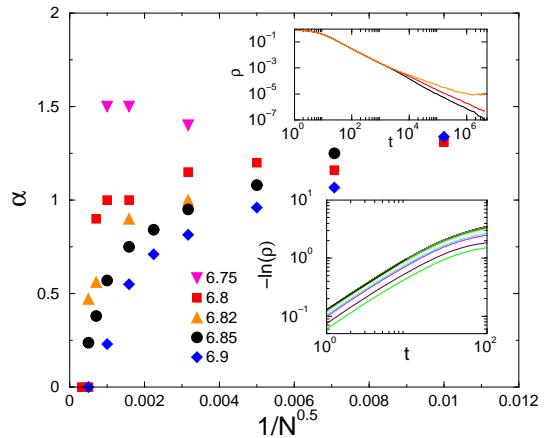}
\caption{\label{fssgwbacp} (Color online)
  Finite-size scaling analysis of the density
  decay exponent for $\lambda=6.75$ (triangles), $\lambda=6.8$
  (boxes), $\lambda=6.82$ (triangles), $\lambda=6.85$ (bullets),
  $\lambda=6.9$ (rhombes) in the CP on weighted BA trees with a
  age-dependent weighting scheme (WBAT-II) with exponent $x=2$.  Top
  inset: $\rho(t)$ for $\lambda=6.82$ ($N=10^6$, $N=4\times10^5$, $N=10^5$
  from top to bottom). Bottom inset: Initial time density near the transition.
  \vspace{0.5cm} }
\end{figure}

Simulations over larger networks hint again towards the disappearance
of the PLs observed in smaller network sizes (see top inset of
Fig.~\ref{fssgwbacp}). In Fig.\ref{fssgwbacp} we plot the value of the
decay exponent for larger systems by performing a power-law regression
in the last two decades of time.  This seems to indicate that as a
function of the network size for $\lambda > 6.8$, the estimated
exponent apparently tends to zero, suggesting the presence of an
active phase in the thermodynamic limit.  On the other hand, for
$\lambda \le 6.8$ the decay exponent does not change too much with $N$
and seems to stay $\alpha \ge 1$ up to the largest sizes $N=4\times
10^6$ we could reach by extensive simulations.

This kind of behavior can again be understood in terms of a smeared
phase transition. Further evidence of this fact comes from the
analysis of the initial decay of the activity density, which again can
be fitted to a stretched exponential of the form
$\rho\propto\exp(t^{-0.87(5)})$ (see bottom inset of
Fig.\ref{fssgwbacp}), in agreement with optimal fluctuation theory in
an infinite dimensional network, Eq.~\eqref{eq:32}. In this case, the
corresponding correlated subspace will be formed by a subset of
connected vertices with alternating small and large degree, and
consequently joined by large weight edges, which would remain active
in the thermodynamic limit giving rise to the characteristic RR
effects observed in simulations. The nature of the RRs in this model
can also be visualized by means of a percolation analysis. 
\begin{figure}[t]
  \includegraphics[height=5cm]{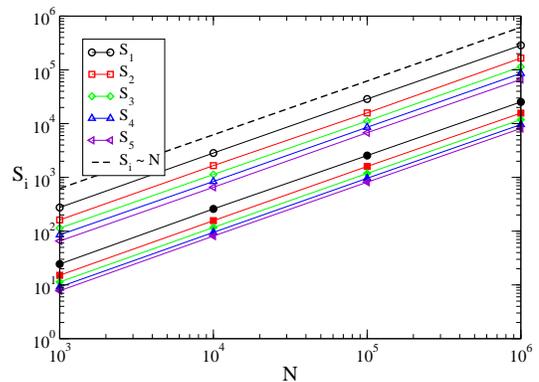}
  \caption{\label{percoII} (Color online)
    Size $S_i$ of the $5$ largest clusters in a
    percolation analysis of the WBAT-II model with $x=2$ for
    $\omega_\mathrm{th} = 100 \omega_\mathrm{min}$ (hollow symbols)
    and $\omega_\mathrm{th} = 1000 \omega_\mathrm{min}$ (full
    symbols), where $ \omega_\mathrm{min}$ is the minimum weight in
    the network. Curves in this last case have been shifted downwards
    for the sake of clarity.}
\end{figure}
In Fig.~\ref{percoII} we plot the average size of the $5$ largest components
in percolation experiments for different values of the weight
threshold  $\omega_\mathrm{th}$. Again, all five scale linearly with
network size, confirming their role as correlated RRs in the
thermodynamic limit.

\begin{figure}[t]
  \includegraphics[height=6cm]{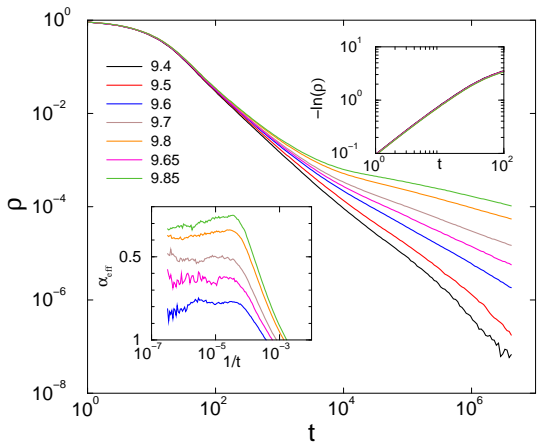}
\caption{\label{modelb-3} (Color online)
  Density decay as a function of time
  $\rho(t)$ for the CP on weighted BA trees with a age-dependent
  weighting scheme (WBAT-II) with exponent $x=3$. Different curves
  correspond to $\lambda=$ 9.4, 9.5, 9.6, 9.65, 9.7, 9.8, 9.85,  
  (from bottom to top). 
  Left inset: The corresponding effective exponents from a local slopes 
  analysis for $\lambda=$ 9.6, 9.65, 9.7, 9.8, 9.85 (from bottom to top). 
  Right inset: initial, stretched exponential behavior.}
\end{figure}

This sort of behavior is not exclusive of the value $x=2$ considered
in the simulations above, but can also be observed for other values of
$x$, see Fig.~\ref{modelb-3}. The initial time dependence for $t< 30$
MCS follows in this case $\rho\propto\exp(t^{-0.87(5)})$ (see right
inset of Fig.~\ref{modelb-3}).

\section{Conclusions}

The heterogeneous and disordered pattern of connections in a complex
network can induce considerable and surprising differences in
dynamical processes running on top of them, as compared with their
outcome on ordered or homogeneous systems.  Building on recent results
concerning the additional effects of intrinsic random, quenched
disorder superimposed over a network structure, in this paper we have
shown that topology-induced disorder can induce rare-region effects on
the contact process.  In particular, we have investigated the effects
of a tree topology, in which no global loops are present in the
network (there is a unique path between any two vertices) and of a
correlated weight pattern assigned to the set of edges. Both elements
have been show to induce a slowing down in dynamics as simple as
diffusion. By means of large scale numerical simulations on tree-like
weighted networks generated using the Barab\'asi-Albert model, we have
shown that a tree-like topology is not by itself enough to induce a
slow dynamics or rare-region effects. However, the combined effects of
a tree structure and a correlated weight pattern are indeed capable of
triggering a very noticeable slow dynamics, which is compatible with
the presence of rare-region effects. The two different weighting
schemes we have considered for small system sizes suggested the
appearance of Griffiths phases, characterized by a power-law dynamics
of the activity and the survival probability above the clean critical
point, with a varying decay exponents.  More accurate simulations
performed on larger network sizes indicate that, instead of true
Griffiths phases, the behavior of the contact process on tree weighted
networks is better explained in terms of a fully smeared transition,
characterized by an initial time decay of the activity density with
the form of an stretched exponential, whose presence can be understood
by optimal fluctuation theory arguments applied on a network with
infinite topological dimension. The smearing of the transition can be
argued to be due to the presence of a correlated subspace of connected
vertices, which are joined by large weight edges, playing the role of
rare-regions and trapping activity at different large time scales.

To sum up, we have shown here that topological disorder is capable to
induce rare-region effects on the contact process on complex
networks. In the particular case we have considered (weighted tree
networks), those rare-region effects translate in the smearing of the
critical phase transition. Our work opens the path to a deeper
investigation of this issue, focusing in particular on discovering
those new topological ingredients which could induce fully-fledged
Griffiths phase originating on topological grounds.

\section*{Acknowledgments}

We thank R. Juh\'asz and I. Kov\'acs for useful discussions, and
C. Castellano and M. A. Mu\~{n}oz for a careful reading of the
manuscript.  R.P.-S. acknowledges financial support from the Spanish
MEC, under project FIS2010-21781-C02-01, and the Junta de
Andaluc\'{\i}a, under project No. P09-FQM4682, as well as additional
support through ICREA Academia, funded by the Generalitat de
Catalunya.  G. \'O. acknowledges support from the Hungarian research
fund OTKA (Grant No. T77629), OSIRIS FP7, HPC-EUROPA2 pr. 228398 and
access to the HUNGRID.

\begin{appendix}

\section{Analytic solution of the HFM equations for WBAT-I model}  
\label{sec:appendix}

In this section we develop the full HMF solution of the WBAT-I model
on uncorrelated networks, sketched in Section
\ref{sec:heter-mean-field}. 

\subsection{Steady state density}

The starting point is Eq.~\eqref{eq:1}, namely
\begin{equation}
  \label{Aeq:1}
  \dot{\rho}_k(t) = - \rho_k(t) + \lambda  [1-\rho_k(t)] \frac{k^{1-\nu}
    \rho(t)}{\av{k^{1-\nu}}},  
\end{equation}
which, in the steady state regime, leads to the non-zero solution for
the partial density of particles in vertices of degree $k$
\begin{equation}
  \label{Aeq:2}
  \rho_k = \frac{\lambda k^{1-\nu} \rho/\av{k^{1-\nu}}}{1+\lambda
    k^{1-\nu} \rho/\av{k^{1-\nu}}}. 
\end{equation}
From this expression, $\rho$, the total density of particles, can be
computed self-consistently, noticing that
\begin{eqnarray}
  \rho &=& \sum_k P(k) \rho_k = \sum_k P(k) \frac{\lambda k^{1-\nu}
    \rho/\av{k^{1-\nu}}}{1+\lambda 
    k^{1-\nu} \rho/\av{k^{1-\nu}}} \nonumber\\
  &=&2m^2 \int_m^\infty \frac{\lambda k^{-2-\nu}
    \rho/\av{k^{1-\nu}}}{1+\lambda 
    k^{1-\nu} \rho/\av{k^{1-\nu}}}\; dk,
  \label{eq:10}
\end{eqnarray}
where in the last expression we have used the degree distribution in
its continuous degree form, $P(k) = 2 m^2 k^{-3}$, and substituted
sums by integrals. Performing the integral in Eq.~\eqref{eq:10} leads
to an equation that can be solved to obtain $\rho$ as a function of
$\lambda$. The form of this equation depends on the value of
$\nu$. Let us consider separately the different cases.

\subsubsection{$\nu=0$}
\label{sec:nu0}

In this case, Eq.~\eqref{eq:10} takes de form
\begin{eqnarray}
  \rho &=& 2m^2 \int_m^\infty \frac{\lambda k^{-2}
    \rho/\av{k}}{1+\lambda 
    k \rho/\av{k}} \; dk\\
  &=&\lambda  \rho  \left[1 - \frac{\lambda  \rho}{2} 
   \ln \left(\frac{2}{\lambda  \rho
   }+1\right)\right],   \label{eq:18}
\end{eqnarray}
where in the last expression we have used $\av{k}=2m$. 
Eq.~\eqref{eq:18} provides an implicit equation for $\rho$
which, in the limit $\rho\to0$, close to the critical point, leads to
the explicit solution
\begin{equation}
  \label{Aeq:19}
  \rho \simeq -\frac{2}{\lambda^2}
    \dfrac{\lambda-1}{\ln\left(\frac{\lambda-1}{\lambda} \right)},
\end{equation}
which corresponds to the asymptotic behavior in Eq.~\eqref{eq:19}.

\subsubsection{$0<\nu<1$}
\label{sec:0nu1}

In this case, performing the integral for $k$ in Eq.~\eqref{eq:10}, we
obtain the self-consistent equation
\begin{equation}
  \label{eq:14}
  \rho= F\left[1, \frac{2}{1-\nu}, 1 +  \frac{2}{1-\nu},
    -\left(\frac{\lambda \rho m^{1-\nu}}{\av{k^{1-\nu}}} \right)^{-1}
  \right],
\end{equation} 
where $F[a,b,c,z]$ is the Gauss hypergeometric function
\cite{abramovitz}. In order to evaluate the critical behavior of the
density in the thermodynamic (infinite network size) lime, we can
invert the previous expression using asymptotic expansion of the Gauss
hypergeometric function \cite{abramovitz}, namely
\begin{eqnarray}
\nonumber
  \lefteqn{F\left[ 1, \alpha, 1+\alpha, -\frac{1}{z}
  \right] =}  \\
&& z^{\alpha}
  \Gamma (\alpha+1) \Gamma \left(1-\alpha\right)  
  + \alpha \sum_{n=1}^\infty (-1)^n \frac{z^n}{n-\alpha}.   \label{Aeq:6}
\end{eqnarray}
For small $\rho$ (small $z$), since $2/(1-\nu)>2$ for $\nu>0$, the most
relevant terms are $z$ and $z^2$. Thus, at leading order, the have
\begin{equation}
  \label{Aeq:7}
  \rho \simeq 
  \frac{2}{1+\nu} \frac{\lambda \rho
    m^{1-\nu}}{\av{k^{1-\nu}}}
  -\frac{1}{\nu} \left(\frac{\lambda \rho
      m^{1-\nu}}{\av{k^{1-\nu}}}\right)^{2} .
\end{equation}
Using now the fact that, in the continuous degree approximation,
$\av{k^{1-\nu}} = 2 m^{1-\nu}/(1+\nu)$, we are led to the expression
\begin{equation}
  \label{Aeq:8}
  \rho \simeq \frac{4 \nu}{(1+\nu)^2 \lambda^2} \left( \lambda-1 \right),
\end{equation}
recovering the asymptotic expression Eq.~\eqref{eq:6}.

\subsubsection{$\nu>1$}

Now, the self-consistent equation for $\rho$ is
\begin{equation}
  \label{eq:15}
  \rho = \lambda \rho \Gamma\left( \frac{2 \nu}{\nu-1} \right) 
  \tilde{F}\left[ 1, \frac{\nu+1}{\nu-1}, 1+  \frac{\nu+1}{\nu-1},
    - \frac{\lambda m^{1-\nu} \rho}{\av{k^{1-\nu}}}  \right],
\end{equation}
where $\tilde{F}[a,b,c,z]$ is the regularized Gauss hypergeometric
function. Using the series expansion for small $z$ \cite{abramovitz}
\begin{equation}
  \label{eq:16}
  \Gamma(\alpha+1) \tilde{F}[1, \alpha, \alpha+1, -z] =
  \alpha \sum_{n=0}^\infty \frac{(-1)^n z^n}{n+\alpha},
\end{equation}
the self-consistent equation at leading order in $\rho$ takes the form
\begin{equation}
  \label{eq:17}
  \rho \simeq \lambda \rho \left[ 1- \frac{\nu-1}{\nu+1} \frac{\lambda
      m^{1-\nu}}{\av{k^{1-\nu}}} \rho \right],
\end{equation}
from where it follows that
\begin{equation}
  \label{eq:9}
  \rho \simeq \frac{2}{(\nu-1)\lambda^2} (\lambda -1),
\end{equation}
from which the scaling form Eq.~\eqref{eq:6} ensues.

\subsection{Density decay at criticality}
\label{sec:density-decay-at}

The decay of the particle density at criticality can be obtained from
Eq.~\eqref{eq:20}, namely
\begin{eqnarray}
  \dot{\rho}(t) &=& \sum_k P(k) \dot{\rho}_k(t) = -
  \frac{\rho(t)}{\av{k}} \sum_k P(k) k \rho_k(t) \nonumber \\
  &=&m \rho(t) \int_m^\infty \frac{\lambda k^{-1-\nu}
    \rho(t)/\av{k^{1-\nu}}}{1+\lambda 
    k^{1-\nu} \rho(t)/\av{k^{1-\nu}}}\; dk,
  \label{eq:11}
\end{eqnarray}
where in the last expression we have performed a quasi-stationary
approximation, substituting $\rho_k(t)$ by the functional form in
Eq.~\eqref{Aeq:8}, and passed to the continuous degree approximation.

\subsubsection{$\nu=0$}

Eq.~\eqref{eq:11} takes the form
\begin{eqnarray}
  \label{eq:12}
  \dot{\rho}(t) &=& m \rho(t) \int_m^\infty \frac{\lambda k^{-1}
    \rho(t)/\av{k}}{1+\lambda 
    k \rho(t)/\av{k}}\; dk \nonumber \\
  &=& -\frac{\rho(t)^2}{2} \ln \left( 1+\frac{2}{\rho(t)}.
  \right)
\end{eqnarray}
This equation can be integrated, yielding
\begin{equation}
  \label{eq:22}
  \mathrm{li}\left(1+\frac{2}{\rho(t)} \right ) \simeq t,
\end{equation}
where $ \mathrm{li}(z)$ is the logarithmic integral function
\cite{abramovitz}.  Using the asymptotic expansion $ \mathrm{li}(z)
\sim z/\ln(z)$ , we obtain the density decay in the large $t$ limit
limit given by Eq.~\eqref{eq:23}.

\subsubsection{$\nu>0$}

In the case $0<\nu<1$, integration of Eq.~\eqref{eq:11} leads to 
\begin{equation}
  \label{eq:26}
  \dot{\rho}(t) \simeq -\rho(t) F\left[1,  \frac{1+\nu}{1-\nu},
    1+\frac{1+\nu}{1-\nu}, - \left(\frac{\lambda \rho
        m^{1-\nu}}{\av{k^{1-\nu}}}\right)^{-1} \right].
\end{equation}
Expanding for small density, using Eq.~\eqref{Aeq:6}, and noticing that
at $(\nu+1)/(1-\nu)>1$, we have at leading order in $\rho$,
\begin{equation}
  \label{eq:27}
  \dot{\rho}(t) \simeq -\frac{1+\nu}{\nu} \frac{\lambda 
        m^{1-\nu}}{\av{k^{1-\nu}}} \rho(t)^2,
\end{equation}
leading to the density decay, in the infinite network size limit,
$\rho(t) \sim t^{-1}$.

An analogous calculation, implying the regularized Gauss
hypergeometric functions, can be performed for $\nu>1$, leading to the
same mean-field behavior for the density decay.

\end{appendix}

\bibliographystyle{apsrev4-1}
\bibliography{griffithstree_3}

\end{document}